%Paper: gr-qc/9307011
%From: nkd@iucaa.ernet.in (Dadhich)
%Date: Mon, 12 Jul 93 15:19:22 GMT

\magnification=\magstep1
\baselineskip=18pt
\input paper

\def\gne #1 #2{\ \vphantom{S}^{\raise-0.5pt\hbox{$\scriptstyle #1$}}_
{\raise0.5pt \hbox{$\scriptstyle #2$}}}

\def\ooo #1 #2{\vphantom{S}^{\raise-0.5pt\hbox{$\scriptstyle #1$}}_
{\raise0.5pt \hbox{$\scriptstyle #2$}}}

% Below find two egs of the use of \gne.  The first one is to just produce
% a greater than or equal to sign.  The second one produces a more
% sophisticated symbol.
% $C \gne > \sim 10.$
% $\ooo  {\ T } {\scriptscriptstyle (m)}$

\voffset=-.5truein
\hsize = 6 true in
\pretolerance=10000
\def\n{\noindent}
\def\m{\medskip}
\def\b{\bigskip}
\def\c{\centerline}
\def\n{\noindent}
\def\s{\smallskip}

\line{\hfil IUCAA-20/93 July'93}

\vskip 2 cm

\c{\mid On Singularity Free Spacetimes--II~:}
\c{\mid Geodesic Completeness}
\vskip 1.5 cm
\c{Naresh Dadhich\footnote{$^1$}{E-mail : naresh@iucaa.ernet.in}
and L. K. Patel\footnote{$^2$}{Permanent Address~: Department of Mathematics,
Gujarat University, Ahmedabad - 380 009, India }}
\vskip 1cm

\c{Inter University Centre for Astronomy and Astrophysics,}
\c{Post Bag 4, Ganeshkhind, Pune - 411 007.}

\vskip .7 cm

\c{\bf Abstract}

\vskip 0.35 cm

We study the geodesics of the singularity free metric considered in the
preceding
Paper I and show that they are complete. This
once again demonstrates the absence of singularity. The
geodesic completeness is established in general without reference
to any particular matter distribution. The metric is globally
hyperbolic and causally stable. The question of inapplicability of
the powerful singularity theorems in this case is discussed.

\vfill
\n PACS numbers : 04.20Jb, 98.80Dr
\eject

\n{\bf I. Introduction}
\m
In paper I [1], we have deduced a general form of the singularity
free spacetime with cylindrical symmetry [2] by effecting a simple
and natural inhomogenis
Friedman--Robertson--Walker (FRW) metric with negative curvature.
This metric can represent a perfect fluid, a fluid with radial heat
flow and massless scalar field as well as two distinct vacuum
solutions. The matter field has acceptable physical behaviour and satisfies
energy
and causality
conditions. In here we wish to examine the nature of the geodesics
of the general metric, not necessarily representing a fluid model.
We shall show that the spacetime is geodesically complete and
hence free of singularities. This will however restrain the range
of free parameters occurring in the metric.

In $\S$II we reiterate some general features of the metric while
$\S$III is devoted to a detailed analysis of the non--spacelike
geodesics to establish that they are complete,i.e. they can
be extended to the arbitrary values of their affine parameter. That
means geodesics never terminate indicating absence of singularity.
Since the spacetime is singularity free, how does it avoid the application
of general results of the powerful singularity theorems [3]?
It turns out that all the energy and causality conditions hold good
but the assumption of existence of compact trapped surfaces fails.
Finally this as well as global hyperbolicity and causal stability
of the metric are discussed in $\S$IV.

The geodesic completeness of the singularity--free radiation spacetime
[4] has been established through a detailed analysis [5]. We shall
follow the same procedure to study this general case.

\bigskip

\noindent {\bf II. The Metric}
\m
Even at the risk of repeatation we shall briefly reiterate the
general features of the metric discussed in detail in Paper I.
this is to make the paper self contained enough so that it can
be read independently. From Paper I, we take the metric for singularity free
spacetime and write

$$\eqalignno{ds^2 = C^{2\alpha} (kt) &C^{2a}(mr) (dt^2 - dr^2) - C^{2\beta}(kt)
C^{2b} (mr) dz^2 - m^{-2} S^2 (mr) C^{2\gamma} (kt)\cr
& C^{2c} (mr) d\phi^2&(1)\cr}$$

\n where $C(x) = cosh x, S(x) = sinh x$. This will represent a perfect
fluid universe with cylindrical symmetry with appropriate constraints
on the parameters $\alpha, \beta, \gamma, a,b,c$ and $m, k$. In fact, it gives
rise to a two parameter family (i.e. of the eight parameters only two
remain free) of exact solutons of Einstein's equations. In the natural
comoving coordinates, the fluid velocity field is given by

$${\bf u} = C^\alpha (kt) C^a (mr) dt\eqno(2)$$

\n which is orthogonal to the spacelike hypersurface $t =$ constant.
The fluid congruence is hence rotation free but it has non-zero expansion
and shear which read as follows~:

$$\theta = {(2\gamma + \beta) \over A} kT (kt)\eqno(3)$$

$$\sigma^2 = {2\over 3A^2} (\gamma - \beta)^2 k^2 T^2 (kt)\eqno(4)$$

\n where $T (kt) = tanh (kt)$ and $A = C^\alpha (kt) C^a (mr).$
It is clear that expansion and shear change their sense at
$t = 0$, $\theta \ooo{\geq} < 0$ as $t \ooo{\geq} < 0$). Both tend to zero as
$t \rightarrow \pm \infty$ or $r \rightarrow \infty$ if $\alpha > 0, a > 0.$

The velocity field (2) is not geodetic and the acceleration vector is

$$\dot u_r = -am S (mr) C^{-a-1} (mr) C^{-\alpha} (kt)
\eqno(5)$$

\n The non-geodetic character of the congruence plays the
crucial role in avoiding formation of singularities. It
means that there exists spacelike {\bf pressure gradient}
that counteracts gravitational attraction to provide a bounce
for the universe when it is in contracting phase for $t < 0$. It
converts contraction to expansion at $t = 0$ without formation of a
singularity. The presence of shear also helps in this process
for not letting the congruence to focus.

The regularity of Weyl and Ricci curvatures and the isotropy
of fluid demand $\alpha + \beta = 1$, $\alpha = \gamma$ and,

\item{(i)} $b = c, a = -b (1 + 2b)^{-1}, k = m (1 + 2b)$
\item{(ii)} $b + c = 1, a = -b (1-b), k = 2m$.

\n These are the only two cases for which the metric (1)
yields a family and singularity free models. For the former there
exists no equation of state in general, for $b = - 1/3$ it gives the
Senovilla radiation universe [4] with $\rho = 3p$ while for the latter
it is always $\rho = p$ representing the stiff fluid.

The density and pressure of the fluid in general read as follows~:

$$\eqalignno{8\pi \rho A^2 = &(a - 3b) m^2 + {a \over b} k^2 +
{(b-a) (3b+a)\over 4b^2} k^2 T^2 (kt)\cr
&+ {1\over 2} ( a - 3b) (b+c -1) m^2 T^2 (mr)&(6)\cr}$$

$$\eqalignno{8\pi p A^2 = &(a+b) m^2 - k^2 + {(b - a) (3b +a) \over
4b^2} k^2 T^2 (kt)\cr
&+ {1 \over 2} (a+b) (b+c-1) T^2 (mr)&(7)\cr}$$

\n where $A = C^\alpha (kt) C^a (mr)$. It can be easily verified
from the above expressions that $\rho = 3p$ for $b = -1/3 = c$ and
$\rho = p$ for $b+c=1$. At a given $t, \rho$ and $p$ are maximum at
$r = 0$ and the overall maximum occurs at $t = 0$ and $r = 0$. The
parameter $m$ or $k$ can be identified with the largest value of energy
density, that can be chosen as large (or small) as one wishes. $\rho$
and $p$ can be made positive everywhere satisfying the strong energy
condition. Since $\rho$ and $p$ are regular and finite everywhere,
it indicates absence of physical (or Ricci) singularity. We have
verified that the metric (1)  is free of Weyl singularity as all Weyl
curvatures are regular and finite and so are the kinematic
parameters.

The metric admits two spacelike killing vectors ${\partial \over
\partial z}$ and ${\partial \over \partial \phi}$ which are
mutually as well as hypersurface orthogonal. The spacetime is cylindrically
symmetric
with $2\pi$ period for the angular coordinate $\phi$. The singularity
at $r = 0$ is the coordinate singularity of cylindrical coordinates.
The $t$ coordinate decreases or increases along every past or future
directed non-spacelike curve, and hence it defines a cosmic time
for the spacetime. The gradient of $t$ always remains timelike which
implies the causal stability of the metric. It is the stronger condition
that includes weaker chronology and causality conditions.

Having seen that the metric (1) is free of the curvature singularities and
also satisfies proper causality conditions, we shall in the next
Section show that the spacetime is geodesically complete. That means it cannot
be extended, ruling out the possibility of hidden singularities. This
is though quite transparent from the form of the metric, we shall however
show it explicitly by studying the geodesics of the metric in detail.

\b

\n{\bf III. Geodesics}
\m
To demonstrate the geodesic completeness of the metric,
we should show that all non-spacelike geodesics can be
extended to arbitrary values of the affine parameter.
For this let us write the geodesics of the metric
(1) following the standard procedure and they would read
as follows~:

$$\eqalignno{~&C^{2\alpha} (kt) C^{2a} (mr) (\dot t^2 - \dot r^2) - L^2
C^{-2\beta} (kt) C^{-2b} (mr)\cr
&- m^2 M^2 S^{-2} (mr) C^{-2\gamma} (kt) C^{-2c} (mr) = \delta,&(8)\cr}$$

$$C^{2\beta} (kt) C^{2b} (mr) \dot z = L,\eqno(9)$$

$$m^{-2} S^2 (mr) C^{2\gamma} (kt) C^{2c} (mr) \dot \phi = M,
\eqno(10)$$

$$\eqalignno{~&\ddot r + am T (mr) (\dot t^2 + \dot r^2) + 2 \alpha k
T (kt) \dot t \dot r\cr
&- M^2 m^3 S^{-3} (mr) C^{-2 (\gamma + \alpha)} (kt) C^{-2 (c + a) + 1}
(mr) (1 + c T^2 (mr))\cr
&- L^2 bm S (mr) C^{-2 (\beta + \alpha)} (kt) C^{-2 (b + a) - 1}
(mr) = 0,&(11)\cr}$$

$$\eqalignno{~&\ddot t + \alpha k T (kt) (\dot t^2 + \dot r^2)
+ 2 am T (mr) \dot t \dot r\cr
&+ \gamma k m^2 M^2 S^{-2} (mr) S (kt) C^{-2 (\gamma + \alpha) -1} (kt)
C^{-2 (c + a)} (mr)\cr
&+\beta k L^2 S (kt) C^{-2 (\beta + \alpha) - 1} (kt) C^{-2 (b + a)}
(mr) = 0.&(12)\cr}$$

\n A dot denotes derivative with respect to the affine parameter,
and $L$ and $M$ are the two constants of motion corresponding to
the two killing vectors representing the conserved $z$ and $\phi$
momenta. The another constant of motion is $\delta$ due to the rest
mass which is one for timelike and zero for null geodesics. Since all
the terms in equations (8--12) are non-singular hence the solutions
to the equations will exist and they will be unique.

We shall now examine the behaviour of first and second derivatives
of the coordinates relative to the affine parameter. We should put finite
bounds on the first derivatives [6] which will imply that the geodesics
are complete. However, the second derivatives should not be singular
to ensure that the field is overall non-singular. Without any loss
of generality we can restrict to the future pointing geodesics.
In this discussion, we donot restrict to fluid distributions,
i.e. the parameters $\alpha, \beta, \gamma, a, b, c, m, k$ are treated
as arbitrary. Following the ref. [5], we shall first consider the
particular geodesics and shall then come to the general case.
\m

\n{\it (a) Fluuid congruence~:} The simplest geodesics are
the ones associated with the fluid congruence. For these we have
$\dot r = \dot \phi = \dot z = 0$. It follows from eq. (2) that
the only geodesic possible in this case has to lie on the axis
$r = 0$ and it is given by

$$\dot t = C^{-\alpha} (kt) \leq 1~~~{\rm for}~~~\alpha \geq 0.$$

\n That means $\alpha$ must be non-negative and then the geodesics
are obviously complete.

\m
\n{\it (b) Axial geodesics~:} For the geodesics lying on the
axis we have $\dot r = 0 = \dot \phi, r = 0$ and so we
write

$$\dot z = \dot L C^{-2\beta} (kt),$$

$$\dot t = C^{-\alpha} (kt) [L^2 C^{-2\beta} (kt) + \delta]^{1/2}
\leq (L^2 + \delta)^{1/2},$$

\n provided $\alpha + \beta \geq 0$. Then the derivatives will
be bounded everywhere and the geodesics will be complete.
Note that the coordinate $t$ attains infinite value only when
so does the affine parameter. This property will be shared by
all the geodesics and will not be referred to henceforth.

\m
\n{\it (c) Radial null geodesics~:} Here $\dot z = 0 = \dot \phi,
\delta = 0$, the first integrals give

$$\dot t = \vert \dot r \vert,~~~\dot t = h C^{-2\alpha} (kt)
C^{-2a} (mr), ~~~h = {\rm const.}$$

\n It is clear that $\vert \dot r \vert = \dot t \leq h$ provided
$\alpha \geq 0, a \geq 0$ and then the geodesics will be complete.
The geodesics with fixed $\phi$ comfortably continue along
$\pi + \phi$ after crossing the axis.

\m
\n{\it (d) Radial timelike geodesics:} Here $\delta = 1$,
$\dot z = 0 = \dot \phi$. Let us parametrise $\dot t$ and $\dot r$
by writing

$$\dot t = C (v) A (r,t), \dot r = S (v) A (r,t), \dot v = B (r,t,v)
A (r,t)$$

\n where $A = C^{-\alpha} (kt) C^{-a} (mr).$ Then

$$B = - [am T (mr) C (v)  + \alpha k T (kt) S (v)].$$

\n We have $a \geq 0, \alpha \geq 0$ and take
$k, m \geq 0$ which will imply $\dot v \leq 0$. We shall consider
the role of the second derivatives in the end. The same reasoning
will apply here to show that geodesics are complete.

\m

\n{\it (e) Null geodesics with zero angular momentum~:} In this
case we have $\dot \phi = 0, \delta = 0$ and as before we write

$$\dot t = C (v) E (r,t), \dot r = S (v) E (r,t),$$

\n and

$$\dot v = E (r,t) F (r,t,v).$$

\n Then we obtain

$$E = \vert L \vert C^{-(\alpha + \beta)} (kt)
C^{- (a + b)} (mr),$$

$$F = - [k (\alpha - \beta) T (kt) S (v)  + m (a - b)
T (mr) C (v)]$$

We shall require $\alpha + \beta \geq 0$, $a + b \geq 0,$
$\alpha \geq \beta$ and $a \geq b$ for $\dot v \leq 0$.

\m
\n{\it (f) Null geodesics on the hypersurfaces $z =$ const.~:}
These are defined by $\dot z = 0, \delta = 0,$ and
as before we have

$$\dot t = C (v) P (r,t), \dot r = S (v) P (r,t)$$

\n and

$$\dot v = P (r,t) D (r,v)$$

\n where

$$P = \vert M \vert m S^{-1} (mr) C^{- (\gamma + \alpha)} (kt)
C^{- (a + c)} (mr),$$

$$D = k (\gamma - \alpha) T (kt) S (v) + m C (v)
[(c -a) T (mr) + T^{-1} (mr)].$$

\n As has been noted earlier that the regularity of the Ricci
tensor requires $\alpha = \gamma$. Hence $D = D (r,v)$ as
assumed.

In this case we can obtain one of the equations of the orbit
that on integration yields

$$C (v) = N^{-1} S (mr) C^{c-a} (mr), N = {\rm const.} > 0.$$

\n Since $C (v) \geq 1,$ the coordinate $r$ is bounded between
the roots of $S (mr) C^{c-a} (mr) = N$. Clearly $\vert v \vert$
is bounded and the geodesics are complete. Among them there could be a
circular geodesic, corresponding to the double root, which will go
around the axis at the fixed radius.

\m
\n{\it (g) General non-spacelike geodesics~:} Now we come to the
general case. From eq. (11) it follows that $\ddot r$ will be
negative for positive $t$ and increasing $r$ provided
$b \leq 0$ (for large $r$ the term with $L^2$ will dominate
over that with $M^2$). With $\ddot r < 0$, the $r$-coordinate
cannot diverge to infinity in a finite proper time. As $r$ decreases
the $\dot t \dot r$ term will dominate over the $(\dot t^2 + \dot r^2)$
term while the $L^2$ term will tend to zero as $r \rightarrow 0$.
In the vicinity of the axis $r = 0$, $\ddot r > 0$ for decreasing
$r$ and $t > 0$. Thus the geodesics cannot collapse quickly enough
into the axis to become singular. It is this feature that really
provides the {\it bounce off} to the universe, turning contraction to
expansion at $t = 0$.

The above arguments will apply to the cases (d) and (e) above and hence
the geodesics will be complete for them. As regards the $t$-coordinate,
$\ddot t$ should be negative for large values of $t$, so that it keeps
its growth in check, not letting it diverge to infinity for finite
values of the affine parameter. From eq. (12), it can be
seen that this will be so even if $\beta \leq 0$ because the
$(\dot t^2 + \dot r^2)$ will always be the dominant term (for
decreasing $r$ the $\dot t \dot r$ term will not be relevant).
$\dot z$ is always regular so long as $t$ and $r$ are regular. $\dot \phi$
diverges as $r \rightarrow 0$ but with $\dot \phi \not= 0$, the geodesics
can never reach $r = 0$ as has been demonstrated in (f) above. In
the neighbourhood of the axis eq. (10) simply represents the
centrifugal effect.

It is straightforward to carry out the similar calculations as done in the
previous cases but they are very combersome and not very enlightening
and hence we will not report them here. It can be verified that the general
geodesics are also complete. Let us now collect together all the restrictions
prescribed by the above considerations on the parameters and they are~:

$$\alpha \geq 0, \alpha + \beta \geq 0, \alpha \geq \beta$$

$$a \geq 0, a + b \geq 0, a \geq b, b \leq 0, k \geq 0, m \geq 0$$

\n and $\alpha = \gamma$ implied by the regularity of the Ricci
tensor. It is interesting to note that $\alpha, \beta$ and $a, b$
satisfy the similar conditions. All this follows purely by requiring
the geodesics to be complete without any reference to the matter
distribution. That is in this framework one can {\it a priori}
ensure the geodesic completeness by adhering to the above conditions on
the parameters. The singularity free
perfect fluid models discussed in Paper I satisfy all these restrictions.

\b
\n{\bf IV. Discussion}
\m
It follows that all the conclusions drawn in Ref. [5] for
the radiation model [4] remain true for this general case as well.
We summarise some of them in the following:

The surfaces $t =$ const. are global cauchy surfaces and hence the
spacetime is globally hyperbolic. This will mean that it is also
causally simple. This is quite obvious from this analysis of the previous
section. Since every $t =$ const. hypersurface is a global cauchy surface
and $t$ defines a time function in the spacetime, it follows that every
non-spacelike curve (need not be geodesic) can be extended to arbitrary
values of its generalised affine parameter as it has to cross all the
cauchy surfaces. That means the spacetime is {\it singularity free}.

It is transparent from the expression for $\rho$ and $p$ that the strong
energy as well as the generic conditions are satisfied. This indicates
that our spacetime satisfies all the energy and causality conditions.
However, it fails to obey the condition of the existence of compact
trapped surfaces, which has been assumed in the singularity theorems.
And this lets it escape the application of the theorems. We shall
now demonstrate that the spacetime does not admit trapped surfaces.

For this we should compute trace of the two null second fundamental
forms [3] and that is given by

$$\eqalign{(g^{ab} X_{ab})_\mp&= {1 \over \sqrt{2}}
C^{-\alpha} (kt) C^{- a + 1} (mr) S^{-1} (mr)\cr
& [ \mp m \mp m (b + c) T^2 (mr) - k (\gamma + \beta)
T (kt) T (mr)]\cr}$$

\n It is clear that $(X^a_a)_- \leq 0$ provided $b + c + 1 \geq 0$
and $(X^a_a)_+ \geq 0$ for $m (1 + b + c) \geq k (\alpha + \beta)$.
Since the traces have opposite signs, hence there are no closed
trapped surfaces.

This means that the outgoing and incoming radial null geodesics are
respectively expanding and contracting everywhere. For existence of a
closed trapped surface, in some region they should all be
contracting, i.e. the trace should be $> 0$ for the both in some
region. It can be verified that the singularity free models of
Paper I satisfy the conditions just deduced. Thus there occur no closed
trapped surfaces in the singularity free family of metrics.

In Paper I (and Ref. [2]), we have deduced a general singularity
free form for the cylindrically symmetric metric. In this paper, we have
demonstrated by analysing its geodesics that the general metric is
{\it really free} of singularities with proper restrictions on the
parameters. This is a general analysis without reference to any matter
distribution. That is the constraints on the parameters deduced above
have always to be respected may what be the energy content of the universe.

The non-existence of closed trapped surfaces can be ensured in advance
so as to escape the application of the singularity theorems [3].
This however does not conflict with the requirements of the energy
as well as the fluid consistency conditions.
That is dropping of this assumption has no adverse physical implications,
makes the theorems inapplicable and provides a singularity free
cosmological model.

One of us (LKP) thanks IUCAA for hospitality.

\vfill\eject

\n{\bf References}
\b
\item{[1]}N. Dadhich, L.K. Patel and R.S. Tikekar, Paper I.
\s
\item{[2]}E. Ruiz and J.M. Senovilla, Phys. Rev. D 45, 1995 (1992).
\s
\item{[3]}S.W. Hawking and G.F.R. Ellis, The Large Scale Structure of
Spacetime (Cambridge University Press, 1973).
\s
\item{[4]}J.M.M. Senovilla, Phys. Rev. Lett. {\bf 64}, 2219
(1990).
\s
\item{[5]}F.J. Chinea, L. Fernandez-Jambrina and J.M.M. Senovilla, Phys.
Rev. D 45, 481 (1992).
\s
\item{[6]}V.I. Arnold, Ordinary Differential Equations (MIT Press,
Cambridge, MA, 1973).
\bye